# Particle Swarm Optimization Approaches for Primary User Emulation Attack Detection and Localization in Cognitive Radio Networks

Walid R. Ghanem[1], Reem E. Mohamed[2], Mona Shokair[3], Moawad I. Dessouky[3]
1 Institute for Digital Communications Friedrich-Alexander University Erlangen-Nürnberg (FAU), Germany
2 Electronics and Communications Engineering Department, Faculty of Engineering, Mansoura University, Mansoura, Egypt,
3 The Department of Electronic and Electrical Communication, Menoufia University, Egypt
Corresponding Author Email: walid.ghanem@fau.de   r.emk.sherif@gmail.com

**Abstract** The primary user emulation attack (PUEA) is one of the common threats in cognitive radio networks (CRNs), in this problem, an attacker mimics the Primary User (PU) signal to deceive other secondary users (SUs) to make them leave the white spaces (free spaces) in the spectrum assigned by the PU. In this paper, the PUEA is detected and localized using the Time-Difference-Of-Arrival (TDOA) localization technique. Particle Swarm Optimization (PSO) algorithms are proposed to solve the cost function of TDOA measurements. The PSO variants are developed by changing the parameters of the standard PSO such as inertia weight and acceleration constants. These approaches are presented and compared with the standard PSO in terms of convergence speed and processing time. This paper presents the first study of designing a PSO algorithm suitable for the localization problem and will be considered as a good guidance for applying the optimization algorithms in wireless positioning techniques. Mean square error (MSE) and cumulative distribution function (CDF) are used as the evaluation metrics to measure the accuracy of the proposed algorithms. Simulation results show that the proposed PSO approaches provide higher accuracy and faster convergence than the standard PSO and the Taylor series estimation (TSE).

## 1. Introduction

The rapidly increasing growth in the wireless technology and the need for adding more new wireless services make the wireless frequency spectrum overcrowded and little space is left for adding new extra applications, a novel technology called "Cognitive Radio" (CR) has been evolved to solve this situation. CR has been proposed to solve the spectrum shortage problem and to improve the efficiency of channel utilization through sharing the resources among the licensed users (Primary Users) and unlicensed users (Secondary Users) [1]. The cognitive radio relies on four basic operations, these operations are spectrum sensing, spectrum management, spectrum sharing and spectrum mobility. The spectrum sensing is the basic and the most important step, in which a Secondary User (SU) can define the white spaces of the spectrum of the Primary Users (PUs) [2]. The SUs must have the ability to differentiate between the PU signal and the other SU signals. Thus, when an SU is using the band and detects the existence of the PU, the SU must leave off the band for him and move to another free spectrum session. Thus, protecting the spectrum sensing process of PU against SU attacks is one of the most important challenges to build a real cognitive radio network (CRN).

Many security problems, encounter the CRN [3], one of these problems called Primary User Emulation Attack (PUEA). PUEA is first presented in [4], where an attacker emulates the PU signal characteristics to deceive the other SUs and steal the free spectrum or corrupt the spectrum sensing process. This attack is performed when the attacker receives the PU signal mimics its features such as transmitting power, modulation, and cyclic prefix to be retransmitted to the other SUs as its own, which prevents PUs from accessing the free bands. The existing techniques for spectrum sensing such as energy detection, matched filter, and cyclostationary feature detection are incapable of defining the attacker due to transmitting a signal with the same power, cyclic prefix and features of the PU signal. Consequently, there is a growing need for supplying active methods to differentiate between legitimate and fake primary transmissions. Therefore, solving this problem is crucial for building a real CRN.

In this paper, we extend our work in [5] that focuses on solving the PUEA on the approved standard IEEE 802.22 Wireless Regional Area Network (WRAN) [6]. The basic goal of the IEEE 802.22 is to utilize the free band in the Digital Television channel (DTV) and to exploit them to provide broadband access to rural areas because the cables media such as coaxial cables increase the system cost. In IEEE 802.22, the PU is a TV tower with a fixed position and located outside the CRN, the CRN consists of a CR base station and a set of CR users (SUs) also with fixed positions and is randomly distributed along the CRN.  The PUEA is detected and localized based on TDOA localization technique. Cooperation is made between the CR users rely on TDOA localization technique. Each SU makes spectrum sensing and sends its recorded measurements to the CR base station, which collects the measurements, and applies the cross-correlation method to extract the TDOA values. After that, these measurements are given to the particle swarm optimization (PSO) algorithms to minimize the cost function and provide an accurate estimation of the unknown transmitter position, which can be a PU (TV tower) or an attacker.

Modified approaches of the PSO are presented based on changing the PSO parameters such as inertia weight and acceleration coefficients. The analytical optimization methods usually rely on intelligent algorithms such as nature-inspired algorithms when network infrastructure expands, which result in more computational time and complexity of these methods. Thus, we propose an algorithm design for CRN that shows low complexity, computational time, memory utilization, and control overhead (parameters), while providing high flexibility and accuracy without wasting the spectrum sensing time compared to the existing analytical optimization methods. This algorithm is based on solving the unconstrained optimization minimization localization problem, where the localization error is optimized for better accuracy in location estimation. It also adapts to CRN network expansion and solves this complex localization problem and finds the optimal solution. This paper compares and analyzes different approaches of PSO in terms of time complexity, convergence speed, and accuracy of the estimated location. The best approach is compared with the standard PSO and the Taylor series estimation method and takes as guidance for solving the localization problems in the wireless positioning system. The contributions of this paper can be summarized as follows:

- Defense against PUEA in CRN based on TDOA localization technique using PSO algorithms. The attacker is localized and compared its position with the location of the PU to decide which PUEA or PU, then compare the estimated position with the location each SU to know which SU is performing the emulating process.
- The comparative study discusses the impact of inertia weight and the acceleration constants on the performance of the PSO algorithm.
- Design and select a PSO approach suitable for the localization process and this can be extended to be used in many wireless localization algorithms.
- Comparison between the standard PSO, Taylor series estimation and the proposed PSO in terms of the location accuracy of the attacker.

The rest of paper is organized as follows, in Section 2, the related work corresponds to the PUEA, literature review about the localization techniques and the optimization algorithms will be presented. The system model, problem formulation and how the attacker is detected will be explained in section 3. Particle swarm optimization algorithms will be studied in Section 4. Section 5 summarizes the detection steps of the PUEA based on PSO algorithms and the metrics parameters are evaluated. Simulation results will be included in Section 6. Finally, conclusions are drawn in Section 7.

## 2. Related Work and Literature Review

PUEA has been discussed in many types of research; most of researchers focused on detecting the attacker. Several detection methods are developed such as energy detection, the detection using *the users' profiles*, authentication methods and location detection methods. The energy detection methods depend on energy measurements that are provided by the CR users. These measurements are collected by the CR base station to detect the presence of the PUEA; this detection method can be easily destroyed by an attacker who is adjusting its transmission power. The author in [7] showed how a PUEA emulated the power of the PU signal to confuse the SUs and a new system model with multiple attackers. They a also show how can the energy detection method is used to defeat this type of attacker, this method cannot be effective if the two channels are similar, so the CRN will confuse again between the signals. The authors in [8] proposed a cooperative spectrum sensing that relies on energy detection in the presence on multiple smart PUEAs and investigated how the performance of cooperative spectrum sensing is influenced by the presence of multiple PUEAs in the CR network and how to mitigate the effect of this attack. The author in [9] proposed an approach based on anomaly behavior detection and collaborated to detect the PUEA in cognitive wireless sensor networks (CWSN). In WSN, nodes usually have their own specific behavior pattern, the network creates a profile for every node, and the node profiles are used to detect anomalies in the behavior such in the case of PUEA. The author in [10] used a channel tap power as a radio frequency fingerprint to identify the PUEA from the PU at the physical layer authentication. The author in [11] proposed an encryption method based on hash message authentication code to authenticate the transmission between the PU and the SU to mitigate PUEA. This method is effective but the modification in the PU transmitter can affect the synchronization between the PU transmitter and PU receivers. It decreases the coverage area of the primary network, and FCC states" no modification to the incumbent system (primary system) should be required to accommodate the opportunistic use of the spectrum by secondary users" so the method that does not require any modification is preferred.

In [4] the authors presented the first method to detect a PUEA relies on location by using the measurements provided by spectrum sensing process and this method is called transmitter verification procedure. The transmitter verification procedure employs a location verification scheme to distinguish incumbent signals from unlicensed signals masquerading as incumbent signals, two alternative techniques were proposed to realize location verification: distance ratio test and distance difference test. More location detection methods can be found in [12] and [13]. In [12], the author used a localization defense model which uses both location and signal characteristics of the signal transmitted to verify the PU signal. The localization process depends on Received Signal Strength (RSS) measurements, the CR users collect the RSS and the RSS peak is identified to estimate the location of the transmitter. The decision about the existence of a PU or an attacker is done by comparing the estimated position with the known position of the TV tower. This method is effective when the size of the network is small; as the network dimensions increases, the localization error increases. Based on the state of the art, the RSS localization methods suffer from the following limitations:

- They are limited to ranges of at most 2 km, so RSS is not suitable for IEEE802.22 because the CR base station range is from 30 km to 100 km.
- The number of cooperating stations is from hundreds to thousands and the cooperation among this large number of users is unpredictable
- These methods are very susceptible to high errors due to the indoor/outdoor environments.

The author in [13] proposed a modified cooperative localization method based on time-difference-of-arrival (TDOA) measurements. The BS collects the TDOA measurements from the SUs and Taylor series estimation (TSE) method is used to estimate the position of the transmitter. The method is suitable for IEEE802.22 and can be applied, but solving the localization process based on Taylor Series (TSE) suffers from the following:

- The need for the suitable initial value that provides sufficient results about the attacker location.
- The high complexity, slow convergence, and low position estimation accuracy of the algorithm.

Localization of unknown transmitter tacked along attention in the last decade and stills a very hot topic for researchers in many fields such as military and civil application [14]. The localization techniques divided into two main categories are mobile based and network-based. In mobile based: the mobile node determines its distance by the signals received from the base station or a global positioning system (GPS) system. GPS based techniques depend on devices report their position which has been measured by timing the signals received by GPS satellites. Obviously, this will not be valid for an attacker device, therefore the GPS localization is not suitable for locating an attacker, but localize only the legitimate nodes (SUs) in the network.

In network-based, it depends on the measured parameters such as received signal strength (RSS), time difference of arrival (TDOA), time of arrival (TOA) and angle of arrival (AOA). In RSS the received signal is measured and then by using an appropriate path loss, these measurements are transformed to distances. This method is very good when the size of the network is small because it does not require additional hardware. However, when the network dimension increases, this method becomes inefficient as explained before. In AOA, the position is measured using the calculation of the direction of arrival at two nodes. This method requires antenna arrays at the receiver and it is dramatically affected by the multipath effect. TOA method is measured depending on the propagation time in one way between the transmitter and the sets of the receiver, then it is used to determine the distance. This measured time values are transformed into a set of circular equations when the positions of all receivers become known, the position of the unknown transmitter will be measured. TDOA is a modification of the TOA, it does not use the absolute time, but it uses the time difference between the arrival times at multiple receivers. TDOA takes advantage of the cross-correlation to measure the difference in the TOA of a transmitted signal at two or more pair of nodes. The minimum number of nodes to locate a transmitter in 2-D is three; the three nodes provide 2 TDOA measurements. TDOA requires a strict synchronization between the SUs and the CR base station. The TDOA provides two hyperbolic curves should intersect at one point. In this paper, the TDOA is chosen as the perfect method to localize the attacker. The authors in [15] present a detailed survey of localization algorithms.

Recently, many optimization algorithms have been developed and applied to different applications such as medical, science and engineering. These algorithms are inspired from the nature such as Particle Swarm Optimization (PSO), Genetic Algorithms (GA), Cuckoo Search algorithm (CS), Firefly Algorithm (FA), Social Spider Algorithm (SSA), Bat Algorithm (BA), Ant Colony Optimization (ACO), Bee Colony optimization (BC), Simulated Annealing (SA), Differential Evolution (DE), Krill Herd (KH), Harmonic Search (HS) and Water Cycle Algorithm (WCA). The PSO is still a very popular and efficient algorithm due to its simplicity, fast convergence, low processing time, high accuracy and low requirement of tuning parameter. In this paper, the PSO optimization algorithm is used to minimize the cost function that was provided by the TDOA measurements to detect the PUEA which are not clarified until now. This paper differs from all previous study of using PSO to solve the localization problem, because of designing and selecting the most suitable version of PSO from its approaches. A brief review of the nature-inspired optimization algorithms is found in [16].

Nature-inspired algorithms have been proposed to solve the localization problem. The author in [17] proposed the using of swarm intelligence to solve localization problem, he suggested using PSO which provides good convergence than simulated annealing and also avoid trapping in local optima. In [18-19] the author compares three different algorithms namely, bat optimization algorithm, modified cuckoo search and firefly optimization algorithm and the results show that Firefly is powerful for using it in localization but using these algorithm needs more memory and the computation complexity will increase. In [20], the author used a modified approach of PSO based on evolution strategy to auto-localization of nodes in static UWB networks; the modified approach increases the accuracy of the estimated location. In this paper, we focus on the PSO due to its advantages that will be discussed in details in Section 4, many modified approaches will be presented rely on changing the parameters of the algorithm and the best approach is selected to be compared with the standard PSO and Taylor series estimation. This paper differs from all previous study in dealing with the localization problem, here we do not only use a single PSO approach, but the basic target to design a good one have all advantages of the PSO and also more suitable for the CRN and this the novelty of this work.

## 3. Problem Formulation and System Model

The system model consists of a PU network with a TV transmitter and TV receivers and a CRN composed of a CR base station and N CR users that are randomly distributed along the network at fixed positions as shown in Figure 1. The CR base station knows the positions of all CR users and the TV tower position. The CRN radius varies from 30 km to 100 km and the PU is located outside the CRN at a specific distance from 30 km to 100 km. The following users are defined as follows:

- **Primary User (PU):** A licensed user who has the higher priority or legacy rights on the usage of a specific part of the spectrum, and is assumed to be at a fixed distance from the TV broadcast tower.
- **Secondary User (SU):** A user who has the lower priority in using this spectrum in a way that it does not cause any interference to the PUs. All CR users are trusted and send trusted information to the CR base station.
- **Primary User Emulation Attack:** The objective of the attacker is not to cause any interference with the PUs, but to forestall spectrum and destroy the spectrum sensing process and therefore preventing the SUs from using the free spectrum when PU not transmitting. It has the ability to receive a PU signal emulates it and retransmits its own signal with the same characteristics of a PU signal to confuse the SUs. The attacker can be classified into the following two types:
  - **Selfish PUEA:** the aim of this attack is to maximize the spectrum usage for him, by taking the free band and prevent other SUs from using it.
  - **Malicious PUEA**: the goal of this attack is to obstruct the spectrum sensing process and prevent the SUs from detecting and using the free bands by emulating or replaying the PU signal.

The detection of the attacker in CRN is done based on cooperative localization algorithm. The location of the unknown transmitter is calculated and then compared with the PU location. If the location is the same as the TV tower, then the PU is transmitting, otherwise, a PUEA is in progress. The attacker can be easily detected if it is located inside the CRN, but if its position is near the TV tower, it will be highly threatening and can be confused with the PU. The objectives of this work are to localize the PUEA inside or outside the CRN with high accuracy to correctly detect the attacker when its location is near to the PU position by reducing the localization positioning error while decreasing the number of needed cooperating users and localization time. The detection of the PUEA model has the following set of assumptions.

- The CRN does not have any information about the attacker or its strategy.
- The PUEA can be close to the PU or can be located inside the CR network.
- The PUEA and the PU have the same radio behavior
- The PUEA can transmit a signal with different or the same characteristic of the PU.

To detect the PUEA, our proposed algorithm is based on based on localization technique that has following three main steps where their procedures will be explained in details in section 5.1:

**Step 1:** Matching the characteristics of a received signal with PU and SU signal characteristics.
**Step 2:** Apply the Cooperative localization technique based on TDOA measurements using PSO.
**Step 3:** Comparing the estimated location with the location of the TV tower and determine which is a PU or PUEA.

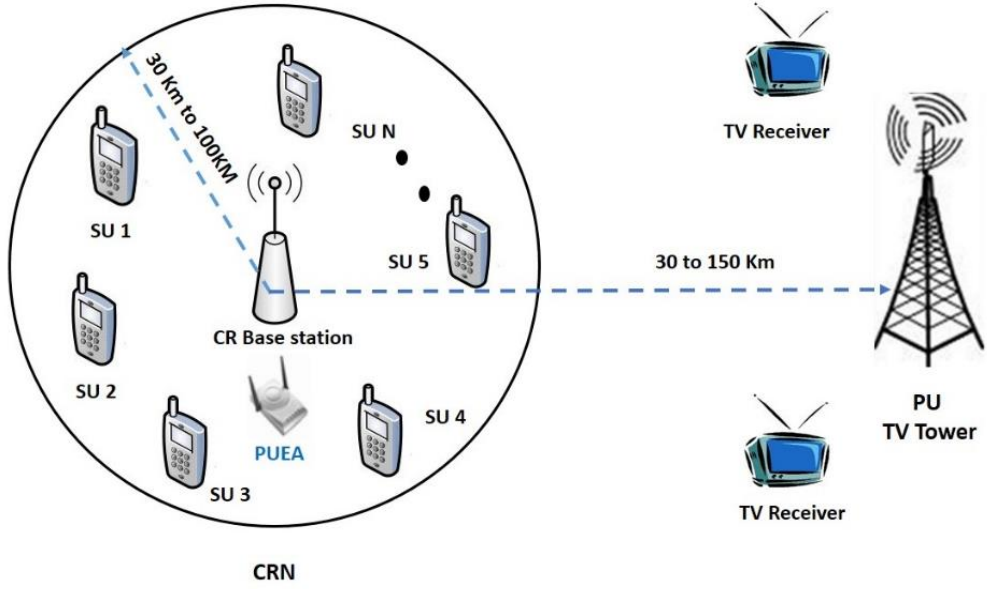

Figure 1 System model

The localization problem can be solved using the TDOA localization technique. All parameters that will be used in the mathematical model of this work are defined as follows:

- Consider $(x, y)$ is the true location of the unknown transmitter, which can be PU or PUEA, $(x_0, y_0)$ is the location of the CR base station.
- It is considered to be the reference station, $(x_i, y_i)$ is the location of the $i^{th}$ SU.
- $d_i$ is the distance between the unknown transmitter and the $i^{th}$ SU, while $d_0$ is the distance between the CR base station and the unknown transmitter.
- $d_{i,0}$ is the actual range difference between the PUEA and the $i^{th}$ SU when the CR base station is considered as a reference node, while $\hat{d}_{i,0}$ is the measured range difference distances.
- $n_i$ is an error which followed a Gaussian distribution with zero mean and variance equal to $\sigma^2$
- $\sigma_i^2$ is the variance of the measurement at the $i^{th}$ SU, and $\sigma_0^2$ is the variance of the measurement at the CR base station
- $(\tilde{x}, \tilde{y})$ is the estimated position of the unknown transmitter
- $B$ is the bandwidth of the TV signal, $SNR_i$ is the Signal to Noise Ratio at the $i^{th}$ SU, while $SNR_0$ is the Signal to Noise Ratio at the CR base station
- $\Delta L_P$ is the used path loss (Hata model) and $h_p$ is the height of the transmitter antenna.

The actual range difference measurement between the PUEA and the $i^{th}$ SU when the CR base station is taken as a reference point can be defined in [21] as given in equation (1)

$$d_{i,0} = d_i - d_0 = \sqrt{(x - x_i)^2 + (y - y_i)^2} - \sqrt{(x - x_0)^2 + (y - y_0)^2} \qquad i = 1,2, \ldots \ldots N \qquad (1)$$

The range difference distance measurements provided by TDOA are corrupted with Gaussian error due to the environment consideration as shown in equation (2)

$$\hat{d}_{i,0} = d_{i,0} + n_i \qquad (2)$$

The position estimation of a given unknown transmitter can be formulated as an optimization problem, involving the minimization of an objective function representing the localization precision. Therefore, each unknown node which can be localized runs stochastic algorithms independently to localize itself by finding its coordinates $(\tilde{x}, \tilde{y})$. The object function or the fitness function for localization problem is given in equation (3)

$$f(\tilde{x}, \tilde{y}) = min \sum_{i=1}^{N} \left( \hat{d}_{i,0} - \sqrt{(\tilde{x} - x_i)^2 + (\tilde{y} - y_i)^2} + \sqrt{(\tilde{x} - x_0)^2 + (\tilde{y} - y_0)^2} \right)^2 \qquad (3)$$

The estimated position of the unknown transmitter $(\tilde{x}, \tilde{y})$ is the value that minimize the fitness function, which is given in equation (3). Assuming $n_i$ is an error, which follows Gaussian distribution with zero mean, and variance that is calculated from the variance of the TDOA $\sigma_i^2$, and the variance of the measurement at the CR base station $\sigma_0^2$ equation (4), the variance of the TDOA is not a fixed value since the SUs have different positions. Therefore, the TDOA variance is modeled in equation (5) given in [13]

$$\sigma^2 = \sigma_i^2 + \sigma_0^2 \quad (4)$$

$$\sigma_i^2 \geq \frac{1}{8\pi^2 \cdot B^2 \cdot SNR_i} \quad (5)$$

The value of $SNR_i$ is given in equation (6)

$$SNR_i = SNR_0 - \Delta L_P(dB) \quad (6)$$

Where $SNR_0$ is the Signal to Noise ratio at the base station, $\Delta L_P(dB)$ is the path loss that is calculated from the Hata model for suburban areas [14], which is path loss model used in IEEE802.22 WRAN and is described in equation (7)

$$\Delta L_P(dB) = [44.9 - 6.55 \cdot Log(h_p)] log(\frac{d_i}{d_0}) \quad (7)$$

## 4. Particle Swarm Optimization

Particle swarm optimization (PSO) is a popular and efficient optimization algorithm, it was developed by Kennedy and Eberhart in 1995 [23], inspired from the social swarm behavior such as birds and fish schools. The PSO searches the space of a fitness function I equation (3) by adjusting the trajectory of individual particles. The particles trace the best location (best solution) in their paths over the course of iterations. In other words, the particles are influenced by their own best locations found as well as the best solution obtained by the whole swarm. The total number of birds is called population size or the swarm size. The goal is to find the global best solution among all current best solutions until the object function no longer improves. The position of the best particle after a fixed number of iterations is the estimated location of the attacker. The PSO has many advantages over other stochastic optimization algorithms, which make PSO more suitable for PUEA detection in CRN. These advantages can be summarized as follows:

- It is easy to implement.
- It requires a few parameters to be adjusted by the user.
- It provides fast convergence, which saves the spectrum sensing process.
- It provides high accuracy.
- In comparison with other heuristics, the PSO is less affected by initial solutions.
- It requires less computational burden compared to other heuristics.
- There are various strategies for reducing the premature convergence.

In PSO, each potential solution to an optimization problem is treated as a fish or a bird. These concepts were modeled using a position vector ($x$) and velocity vectors ($v$) of length $D$, where $D$ indicates the dimension or the number of variables of the problem, in this problem, the attacker is localized in 2D space, therefore this value equals to 2. At each iteration number ($t+1$), the position and velocity of the $j^{th}$ particle from the total swarm particles ($K$) are updated according to the equations (8.a-8.b) in [22]

$$v_j^{t+1} = v_j^t + c_1 r_1 (Pbest_j - x_j^t) + c_2 r_2 (gbest_j - x_j^t) \qquad j = 1, 2, \ldots, K \quad (8.a)$$
$$x_j^{t+1} = x_j^t + v_j^{t+1} \quad (8.b)$$

Where $c_1$ is the cognitive coefficient that controls the influence of the individual memory of good solutions found, $c_2$ is the social factor which controls the extent to which a particle's motion is influenced by the best solution found by the whole swarm, $r_1$, $r_2$ are two random numbers between 0 and 1 which are used to give PSO more randomized search ability, and $Pbest$, $gbest$ are two variables to store the best solutions obtained so far by each particle and the whole swarm, respectively. The efficient optimization algorithm must balance between two important factors, first, the global search which make the algorithm to search more new areas and local search which is used to fine tune the current area. Therefore, the PSO parameters are very important for an optimization problem to succeed. Therefore, Shi and Eberhart proposed a PSO in [23] that depends on inertia weight, and the new position and velocity. It can be calculated from equations (9.a-9.b)

$$v_j^{t+1} = w v_j^t + c_1 r_1 (Pbest_j - x_j^t) + c_2 r_2 (gbest_j - x_j^t) \quad (9.a)$$
$$x_j^{t+1} = x_j^t + v_j^{t+1} \quad (9.b)$$

The following three terms exist in velocity modulation:

- The first part called 'inertia' or 'habit' and describe the tendency of a particle to keep in the same direction during the transverse.
- The second term is called self-knowledge and responsible for moving the particle to its own best experience scaled by a random weight $c_1 r_1$.
- The third part called social knowledge and this responsible for moving the particle to the best experience among particles in the swarm scaled by $c_2 r_2$.

Thus, there are three main coefficients as follows:

- The inertia weight $w$, which is used to control the effect of the previous history of velocities on the current velocity and is responsible for controlling the PSO stability. A large $w$ facilities the global search or exploration, whilst a small *w* improves the local search. Accordingly, a suitable value of *w* makes a balance between the global and the local search, different shapes of inertia weights will be discussed in this paper.
- The two acceleration constant $c_1$ and $c_2$, as the value of $c_1$ increases, it enhances the particle attraction towards *Pbest* and decreases the attraction toward the *gbest*, also increases in $c_2$ enhances the attraction toward *gbest*.

In general, for the PSO parameters setting, this paper uses PSO with fixed parameters, which is the most common approach in previous studies, and PSO variants with dynamic inertia weights.

**4.1 Inertia Weights for PSO**

The Standard PSO (SPSO) suffers from two main disadvantages trapping in local minima and slow convergence rate in solving optimization problems. The dynamic tuning of PSO parameters is a very good way to give the particles different behaviors as the algorithm proceeds, also modified versions of *w* are used to increase the performance of PSO. The inertia weight plays a critical rule to balance the exploration and exploitation, a large value can facilitate a global search, while a small value provides local search. A static value of this parameter does not give a fast convergence or high accuracy. The choice of the best strategy differs from an optimization problem to another, which adds difficulty to the optimization problem. The main objective of our work is to choose and develop the best strategy related to the localization problem that is related to a security problem. Different inertia weights strategies will be used as follows:
- Constant inertia weight: The inertia weight remains fixed during every iteration number (*t*) from the total number of iteration (*T*). For *w* >1.2 the PSO performs weak exploration and *w* <0.8 the PSO will trap in local in local optima, in this paper this value is suggested to be *w*=0.9.
- Linear time-varying inertia weight: The inertia weight varies linearly with the iteration number.
- Random inertia weight: A random value of inertia weights is used to enable the PSO to track the global optima.

Table 1 summarizes the different strategies that will be examined. The parameters that will be used in Table 1 are defined as follows. $w(t)$ is the value of the inertia weight at *t* iterations, $w_{max}$ is the maximum value of the inertia weight, $w_{min}$ is the minimum value of the inertia weight, $rand()$ is a random value between 0,1.

**Table 1** Inertia weights strategies

| Label | Name of inertia Weight | Formula of inertia weight | Ref |
|---|---|---|---|
| W0 | Constant inertia | $w(t) = 0.9$ | [23] |
| W1 | Linear decreasing | $w(t) = w_{max} - \left(\frac{(w_{max} - w_{min})}{T}\right)$ <br> $w_{max}=0.9, w_{min} = 0.4$ | [24] |
| W2 | Random inertia weight | $w(t) = 0.5 + \frac{rand()}{2}$ | [25] |
| W3 | Simulated Annealing | $W(t) = w_{min} + (w_{max} - w_{min}) \times (0.95)^{t-1}$ <br> $w_{max}=0.9, w_{min} = 0.4$ | [26] |
| W4 | Logarithm Decreasing inertia weight | $W(t) = w_{max} + (w_{min} - w_{max}) \times log_{10}(\alpha + \frac{10t}{T})$ <br> $w_{max}=0.9, w_{min} = 0.4$ | [27] |
| W5 | Oscillating inertia weight | $W(t) = \frac{w_{min} + w_{max}}{2} + \frac{w_{min} - w_{max}}{2} \cos(\frac{2pi}{T})$ <br> $w_{max}=0.9, w_{min} = 0.4$ | [28] |
| W6 | Natural Exponent inertia weight | $W(t) = w_{min} + (w_{max} - w_{min})e^{-[\frac{t}{T/10}]}$ <br> $w_{max}=0.9, w_{min} = 0.4$ | [27] |
| W7 | Chaotic inertia weight | $W(t) = (w_{max} - w_{min}) \times \frac{T-t}{T} + w_{min} \times c$ <br> c=4c(1-c), c=0.3 | [29] |
| W8 | Nonlinear decreasing inertia weight | $W(t) = [\frac{(T-1)^n}{(T)^n}](w_{max} - w_{min}) + w_{min}$ <br> $w_{max}=0.9, w_{min} = 0.4, n=0.7$ | [30] |
| W9 | Decline curve with Sugeno function | $W(t) = \frac{1-\beta}{1-s\beta}, \beta = \frac{t}{T}, s= -0.7$ | [30] |
| W10 | Nonlinear increasing inertia weight | $W(t) = w_{initial} \times u^t$<br>$w_{initial}=0.1$, u=1.00002 | [32] |
| W11 | Exponential Inertia Weight | $W(t) = w_{initial} \times e^{-a[\frac{t}{T}]^b}$<br>$w_{initial}=0.4$, $a=2$ and $b=1.5$ | [33] |
| W12 | Chaotic random inertia weight | $W(t) = \frac{rand()}{2} + \frac{c}{2}$<br>C=4*c(1-c), c=0.3 | [29] |

**4.2 Different strategies for acceleration constants**

The acceleration coefficients represent the weighting of the stochastic acceleration terms that pull particles toward *pbest* and *gbest* [34]. If the value of these constants is too large, the particles move abruptly and the risk of being trapped in false optima increases, conversely, if their value is too small, the particles move too slowly, the computation effort increases and the algorithm could not converge. Therefore, a proper control of these components is very important. On the other hand, the relative value of this two acceleration is critical and affect the algorithm's behavior, when the value of $c_1$ increases, it enhances particle attraction towards *pbest* and decreases the attraction toward the *gbest*, also the increase in $c_2$ enhances the attraction towards *gbest*. Three different approaches will be applied and introduced to enchase the original PSO in parallel with using the different inertia weight given by table 1. In this paper, three acceleration constants strategies will be presented namely, SPSO, modified PSO (MPSO) and improved PSO (IPSO). Each strategy will be used side by side with changing the inertia weights that are provided in table 1. For example, for fixed acceleration constants with label A1 and changing the inertia weight by using weights from label W1 to W12, therefore we obtain 12 different approaches namely PSO1, PSO2, to PSO12. Table 3 summarizes all approaches that will be compared in the simulation. Note that the constants related to every inertia weight strategies have come after more than an attempt to get the best results. Table 3 summarizes all PSO approaches that that will be compared with the original particle swarm algorithm (PSO). This table is developed by mixing Table 1 and Table 2.

**Table 2** Acceleration constants strategies

| Algorithm | Label | Updating formula $c_1$ | $c_2$ | Ref |
|---|---|---|---|---|
| PSO | A1 | 2 | 2 | [22] |
| MPSO | A2 | $\left(-\frac{2.05}{T}\right)t + 2.55$ | $\left(\frac{1}{T}\right)t + 1.25$ | [35] |
| IPSO | A3 | $25 + 2(t/T)^2 - 2(\frac{2t}{T})$ | $0.5 - 2\left(\frac{t}{T}\right)^2 + 2(\frac{2t}{T})$ | [36] |

**Table 3** Summary of various PSO modifications that described in the paper

| PSO Name | W label | $c_1, c_2$ Label | PSO Name | W label | $c_1, c_2$ Label | PSO Name | W label | $c_1, c_2$ Label |
|---|---|---|---|---|---|---|---|---|
| **PSO** | W0 | A1 | **MPSO** | W0 | A2 | **IPSO** | W0 | A3 |
| **PSO1** | W1 | A1 | **MPSO1** | W1 | A2 | **IPSO1** | W1 | A3 |
| **PSO2** | W2 | A1 | **MPSO2** | W2 | A2 | **IPSO2** | W2 | A3 |
| **PSO3** | W3 | A1 | **MPSO3** | W3 | A2 | **IPSO3** | W3 | A3 |
| **PSO4** | W4 | A1 | **MPSO4** | W4 | A2 | **IPSO4** | W4 | A3 |
| **PSO5** | W5 | A1 | **MPSO5** | W5 | A2 | **IPSO5** | W5 | A3 |
| **PSO6** | W6 | A1 | **MPSO6** | W6 | A2 | **IPSO6** | W6 | A3 |
| **PSO7** | W7 | A1 | **MPSO7** | W7 | A2 | **IPSO7** | W7 | A3 |
| **PSO8** | W8 | A1 | **MPSO8** | W8 | A2 | **IPSO8** | W8 | A3 |
| **PSO9** | W9 | A1 | **MPSO9** | W9 | A2 | **IPSO9** | W9 | A3 |
| **PSO10** | W10 | A1 | **MPSO10** | W10 | A2 | **IPSO10** | W10 | A3 |
| **PSO11** | W11 | A1 | **MPSO11** | W11 | A2 | **IPSO11** | W11 | A3 |
| **PSO12** | W12 | A1 | **MPSO12** | W12 | A2 | **IPSO12** | W12 | A3 |

## 5. PUEA Detection and Metric Parameter Measurements
In this section, the detection of the attacker depends on PSO algorithms is summarized and how the performance evaluation of the localization algorithms will be measured.

### 5.1 PUEA Detection and Localization Based on PSO
The following steps summarize the localization procedures of the PUEA in cognitive radio networks based on PSO algorithms and Fig 2 gives the flowchart of the PSO detection method
- Deploy 100 SUs randomly in the 30km×30km area of the CRN field and the CR base station at the center.
- Each SU in the CRN makes spectrum sensing process, records and sends the signal to the CR base station.
- The CR base station receives signals from each SU, and applies cross-correlation between every received signal form every SU and its own received signal to produce the TDOA values, by multiply these values with the speed of light, it is transformed into the range distances differences.
- The range distances are calculated from the cost function that is given in equation (5)
- The PSO approaches are then used to minimize equation (5), the position of the bird (particle) which minimizes this equation is the location of the suspected node.
- The location of the suspected node is then compared with the position of the PU to determine if the unknown transmitter is PU or PUEA and then compared with the positions of the SUs to determine which SU is performing the attack.

### 5.2 Metric Parameter for Performance Evaluation
This subsection depicts the evaluation methods for the PUEA localization algorithms based on the PSO, in order to compare the effectiveness of the positioning techniques; we evaluate the accuracy of the proposed algorithms in terms of mean square error (MSE) and cumulative distribution function (CDF). The algorithm that has the higher accuracy is more effective.

#### 5.2.1 Mean Square Error
Mean Square error (MSE) is suitable for measuring the performance of a positioning techniques, The MSE is calculated as follows, suppose the position of the PUEA is estimated n times, such that a population of coordination is calculated $(X_1, Y_2), \ldots, (X_n, Y_n)$, therefore the mean of the calculated coordinates is calculated from equation(10) where the actual coordinates of the attacker is $(x, y)$

$$\overline{x} = \frac{1}{n}\sum_{i=1}^{n} X_i \ , \ \overline{y} = \frac{1}{n}\sum_{i=1}^{n} Y_i \qquad (10)$$

The mean square error of the estimated position is given by equation (11)

$$MSE(\overline{x}, \overline{y}) = E[(\overline{x} - x)^2 + (\overline{y} - y)^2] \qquad (11)$$

#### 5.2.2 Cumulative Distribution Function
Cumulative distribution function (CDF) describes the probability that a real-valued random variable Z with a given probability distribution will be found at a value less than *z* as formulated in equation (12)

$$F_z(z) = P(Z \leq z) \qquad (12)$$

Where Z is a numerical random variable, *F (z)* is called the cumulative disruption function of variable Z and $P(Z \leq z)$ denotes the probability distribution that is found at a value less than z. It can be regarded as the proportion of the population coordination whose value is less than *z*, the CDF is increasing function from 0 to 1. The CDF describe the performance with how many meters of resolution with a certain probability, therefore *z* is the variable of the error distance in meter and $P(Z \leq z)$ represents the probability of the algorithms within *z* meters of error distance.

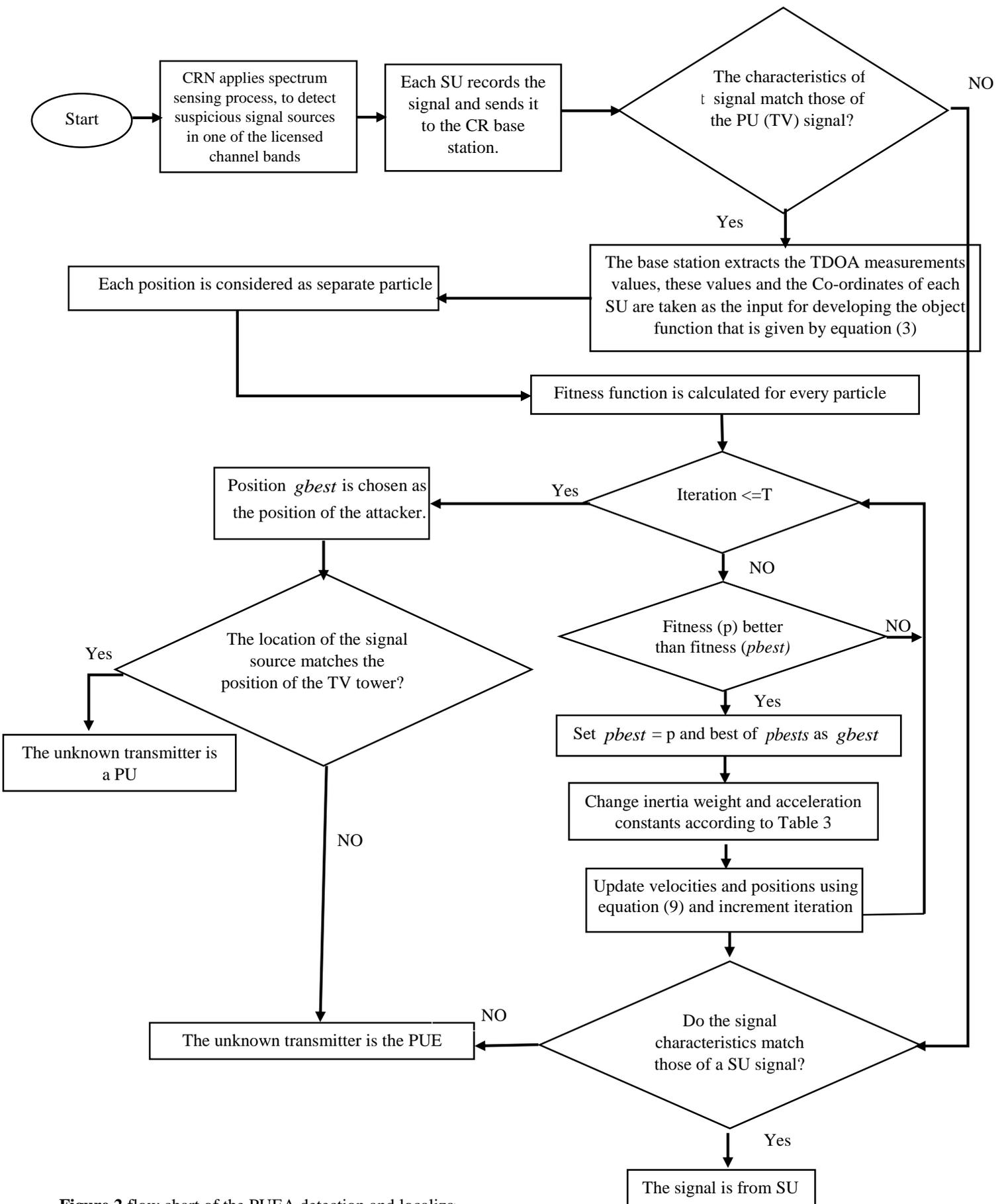

**Figure 2** flow chart of the PUEA detection and localiza based on PSO

## 6. Simulation Results and Discussion

In this section, the effectiveness of each method will be examined, they are compared through simulation in MATLAB program on a laptop of 6 GB memory and 1.7 GHz core i5 CPU to evaluate the performance of proposed algorithms. The simulation area is 30 km ×30 km from the CR base station at the origin (0, 0) and 100 SUs randomly distributed within the boundary of the CR network. The PUEA is located inside the network at location (8000m, 1000m) or outside the network. The performance is measured using the MSE and CDF by running the Monte Carlo 1000 times and averaging the results. The simulation Parameters of the CRN are shown in Table 4. The PSO approaches are used to minimize the fitness function that is given by equation (3), the output of the PSO algorithms are two-fold, first, the position of the particle, which is the estimated position of the PUEA, second the fitness value at this position.

**Table 4** Simulation Parameters of the CRN system model

| Parameter description | Values |
|---|---|
| Area of the CRN | 30km×30km |
| $N$ | 10:100 |
| PUEA antenna height | 1.5m |
| TV tower distance location | 30km:100km Km |
| Bandwidth of TV channel | 6MHz |
| Channel path loss model | Hata model |
| Population size of the PSO ($K$) | 40 |

In order to compare the convergence performance of all algorithms, the number of iterations required to reach a minimum value of the fitness function is calculated. From the extensive simulations, the convergence rates are shown below in Fig 3(a, b, c, d, e, and f). The horizontal axis represents the number of iteration and the vertical axis shows the logarithmic values of the fitness that is described by equation (3) (log (fitness value)). All modified algorithms are compared with the original PSO in each figure. It can be seen from the pictures that the convergence rate of the modified approaches is obviously faster the original PSO. The adding of the inertia weight and the dynamic change of the acceleration coefficients has a great effect on the convergence rate of the PSO. By looking at the convergence curves more closely in Fig 3(a, b, c, d, e, and f), we conclude that PSO10, PSO12, MPSO11, MPSO12, MPSO10, IPSO11 and IPSO12 obtain better results compared with the other PSO algorithms. These seven approaches are chosen to be compared again to each other for accurate comparison.

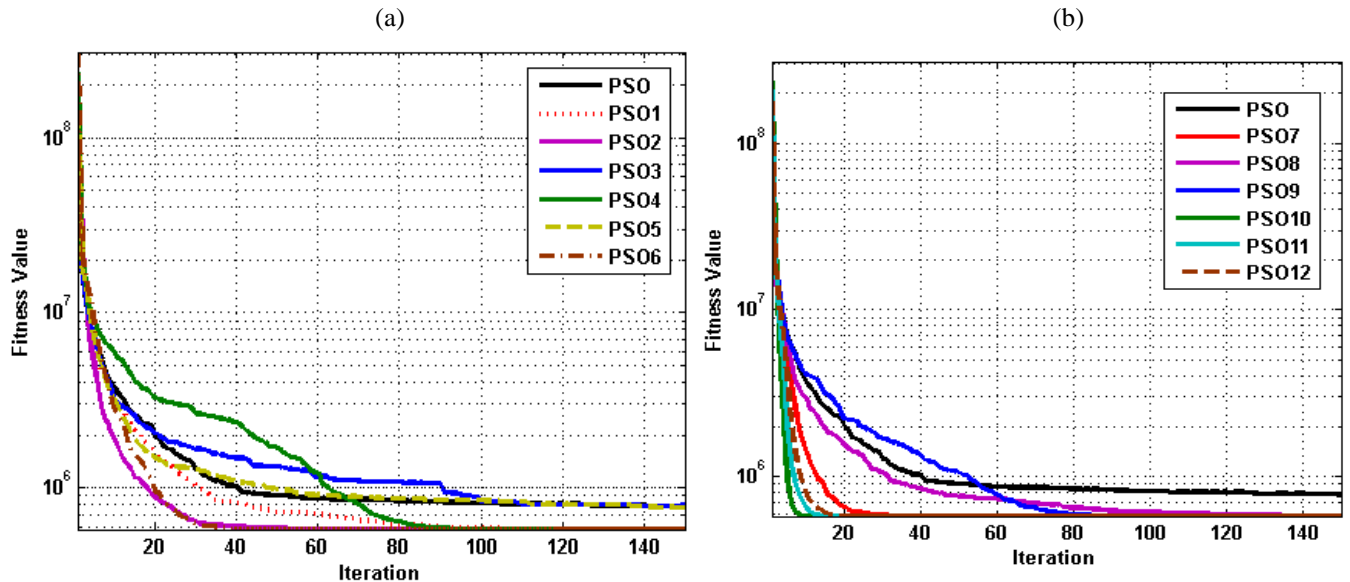

(a)    (b)

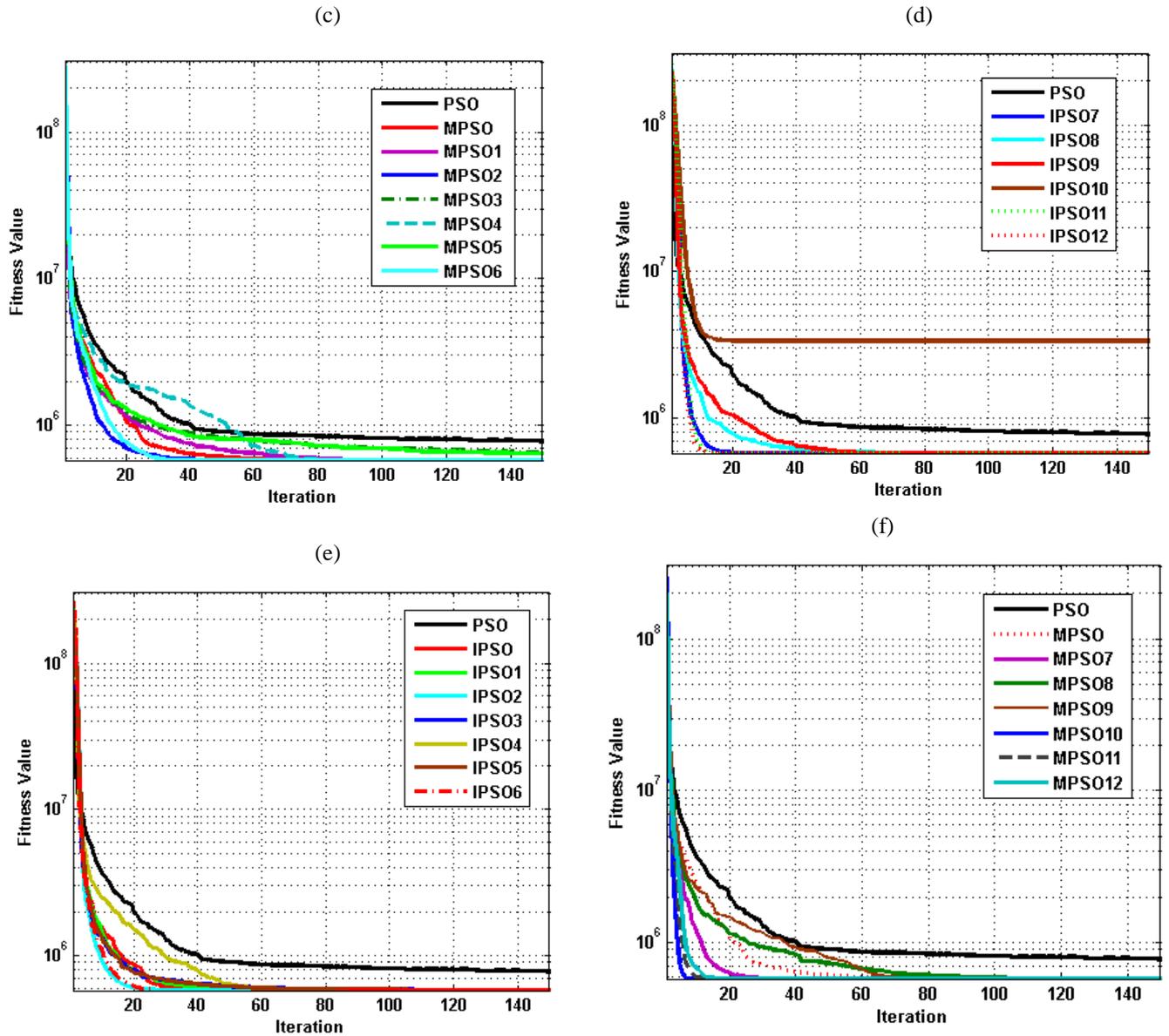

**Figure 3** the impact of the iteration number on the fitness value for (a, b) PSO with inertia weights with labels from (a) W1 to W6, and (b) W7 to W12, (c, d) MPSO with inertia weights (c) W1 to W6, and (d) W7 to W12, and (e, f) IPSO with inertia weights (e) W1 to W6, and (f) W7 to W12

Table 5 gives the approximate convergence time in seconds for each algorithm and the approximate number of iteration required for each algorithm until convergence, these results are compatible with Fig 3(a , b, c, d, and f). The original PSO has the large convergence time and requires a large number of iterations until convergence. All approaches that have minimal processing are shaded with the shaded area in table 3. The best 7 approaches are selected from the table 5 according to the low processing time and compared again. A careful consideration of this table shows that the original PSO gives about 1.315 seconds to convergence while, PSO10, PSO12, MPSO10, MPSO11, MPSO12, IPSO11, IPSO12 take only 0.091 s, 0.108 s, 0.073 s, 0.090 s, 0.082 s, 0.089 s, and 0.090 s. The low processing time is very important to decrease the spectrum sensing time for the CRN.

**Table 5** the Convergence time and the iteration number of the PSO approaches

| S.N | PSO version | Convergence (iteration) | Convergence time | S.N | PSO version | Convergence (iteration) | Convergence time |
|---|---|---|---|---|---|---|---|
| 1 | PSO | 100 | 1.315s | 21 | MPSO7 | 20 | 0.272 s |
| 2 | PSO1 | 90 | 1.076s | 22 | MPSO8 | 40 | 0.380 s |
| 3 | PSO2 | 28 | 0.278 s | 23 | MPSO9 | 70 | 0.621 s |
| 4 | PSO3 | 60 | 0.572s | 24 | MPSO10 | 9 | 0.073 s |
| 5 | PSO4 | 80 | 1.044 s | 25 | MPSO11 | 10 | 0.090 s |
| 6 | PSO5 | 50 | 0.492 s | 26 | MPSO12 | 10 | 0.082 s |
| 7 | PSO6 | 40 | 0.404 s | 27 | IPSO | 25 | 0.156 s |
| 8 | PSO7 | 20 | 0.192 s | 28 | IPSO1 | 20 | 0.312 s |
| 9 | PSO8 | 100 | 1.275 s | 29 | IPSO2 | 20 | 0.205 s |
| 10 | PSO9 | 80 | 0.812 s | 30 | IPSO3 | 20 | 0.191 s |
| 11 | PSO10 | 19 | 0.22 | 31 | IPSO4 | 50 | 0.488 s |
| 12 | PSO11 | 20 | 0.208 s | 32 | IPSO5 | 20 | 0.194 s |
| 13 | PSO12 | 20 | 0.208 s | 33 | IPSO6 | 20 | 0.203 s |
| 14 | MPSO | 76 | 1.077 s | 34 | IPSO7 | 15 | 0.176 s |
| 15 | MPSO1 | 40 | 0.390 s | 35 | IPSO8 | 25 | 0.248 s |
| 16 | MPSO2 | 20 | 0.256 s | 36 | IPSO9 | 20 | 0.204 s |
| 17 | MPSO3 | 20 | 0.211 s | 37 | IPSO10 | 200 | 2.306s |
| 18 | MPSO4 | 60 | 0.559 s | 38 | IPSO11 | 10 | 0.089 s |
| 19 | MPSO5 | 20 | 0.204 s | 39 | IPSO12 | 10 | 0.090 s |
| 20 | MPSO6 | 25 | 0.241 s | | | | |

Obviously, the comparison of the convergence curves is just one way of presenting results, another way is to compare the MSE for each algorithm. The best 7 approaches were obtained from table 5 and compared to each other. Figure 4 gives the effect of the number of iteration of the MSE of PUEA location. The results show that the inertia weight and the acceleration constants have a great effect on the convergence of the PSO. By looking at this figure more closely, we show that MPSO11, MPSO10 and IPSO11 convergence faster than all approaches. The original PSO fails to fine tune the result and trapped in local optima and also needs many numbers of iteration to convergence. At the other side, all approaches need a small number of iteration therefore, the processing time will be small. Thanks to the inertia weights strategies and the dynamic change of the acceleration constants. The results given in Figure 4 are drawn more clearly in Table 6 that summarizes the results that are obtained in Figure 4. The best 7 approaches are compared with each other's, by increasing the iteration value the MSE decreases and this is an expected behavior. The original PSO needs about 140 iterations to convergence and the MSE is about 55 meters, while PSO10, MPSO10, MPSO11 and IPSO11 need only 10 iterations to provide MSE equals to only 8 meters.

**Table 6** The impact of changing the iteration number on the MSE (meters) for the best 7 PSO approaches as

| Iteration | PSO | PSO10 | PSO12 | MPSO10 | MPSO11 | MPSO12 | IPSO11 | IPSO12 |
|---|---|---|---|---|---|---|---|---|
| 1 | 1372.9 | 539.5 | 1432.1 | 1583.4 | 1445.5 | 1412.8 | 1465.8 | 1454.0 |
| 2 | 614.0 | 596.5 | 630.8 | 557.8 | 589.5 | 557.3 | 595.7 | 624.8 |
| 3 | 362.8 | 278.1 | 429.6 | 248.2 | 228.9 | 347.0 | 258.9 | 392.2 |
| 4 | 324.3 | 131.8 | 313.3 | 116.1 | 97.9 | 246.0 | 104.7 | 282.1 |
| 5 | 202.9 | 68.7 | 216.9 | 52.2 | 39.5 | 170.2 | 46.4 | 174.2 |
| 10 | 156.1 | 8.4 | 54.7 | 8.2 | 8.2 | 34.8 | 8.2 | 32.8 |
| 15 | 130.5 | 7.5 | 18.1 | 7.5 | 7.5 | 11.3 | 7.5 | 11.0 |
| 20 | 120.4 | 7.7 | 9.7 | 7.7 | 7.7 | 8.1 | 7.7 | 8.2 |
| 25 | 112.8 | 7.8 | 8.1 | 7.8 | 7.8 | 7.8 | 7.8 | 7.8 |
| 30 | 92.7 | 7.8 | 7.9 | 7.8 | 7.8 | 7.9 | 7.8 | 7.8 |
| 40 | 84.1 | 8.1 | 8.1 | 8.1 | 8.1 | 8.1 | 8.2 | 8.1 |
| 50 | 74.0 | 8.0 | 8.0 | 8.0 | 8.0 | 8.0 | 8.0 | 8.0 |
| 60 | 68.8 | 7.7 | 7.7 | 7.7 | 7.7 | 7.7 | 8.0 | 7.7 |
| 70 | 71.9 | 7.8 | 7.8 | 7.8 | 7.8 | 7.8 | 8.3 | 7.8 |
| 80 | 63.4 | 7.5 | 7.5 | 7.5 | 7.5 | 7.5 | 7.9 | 7.5 |
| 90 | 64.7 | 8.1 | 8.1 | 8.1 | 8.1 | 8.1 | 8.7 | 8.1 |
| 100 | 60.8 | 7.5 | 7.5 | 7.5 | 7.5 | 7.5 | 7.6 | 7.5 |
| 110 | 60.9 | 7.6 | 7.6 | 7.6 | 7.6 | 7.6 | 7.7 | 7.6 |
| 120 | 61.6 | 7.5 | 7.5 | 7.5 | 7.5 | 7.5 | 7.5 | 7.5 |
| 130 | 62.5 | 7.7 | 7.7 | 7.7 | 7.7 | 7.7 | 7.7 | 7.7 |
| 140 | 55.9 | 8.1 | 8.1 | 8.1 | 8.1 | 8.1 | 8.1 | 8.1 |
| 150 | 55.9 | 7.6 | 7.6 | 7.6 | 7.6 | 7.6 | 7.6 | 7.6 |

Figure 5 shows the comparison between the CDF of the PUEA localization error for the three algorithms, it is seen that the performance of the MPSO11 is higher than that of the original PSO and TSE method in [10]. The proposed MPSO11 is used only 10 iterations to provide this results and PSO has used about 150 iterations when the swarm size is kept 40. We can observe that the proposed approach enhanced the localization accuracy. At CDF equals to 0.5 the distance error corresponds to about 75, 100 and 110 for the MPSO11, PSO, and the TSE methods. Therefore the MPSO11 improves the localization error distance to about 35 meters compared to the TSE method. Thanks to the exponential inertia weight strategy that has the label W11 in Table 1 that has the gratitude for improving the local search of the original PSO.

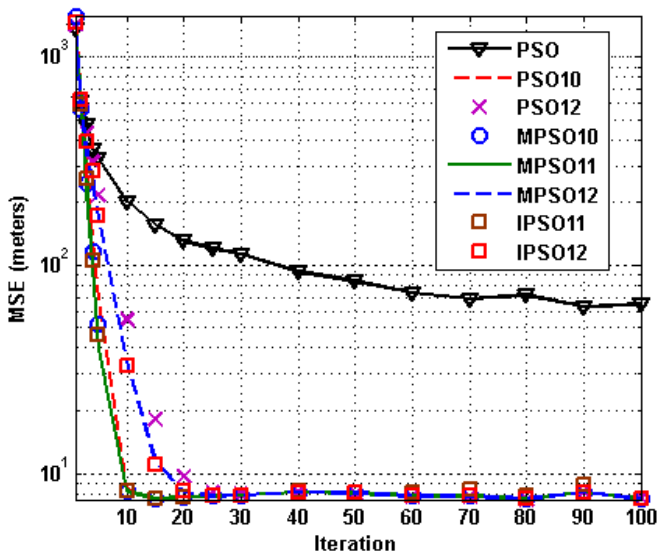

**Figure 4** the impact of the number of iterations on the MSE (meters) of the PUEA position

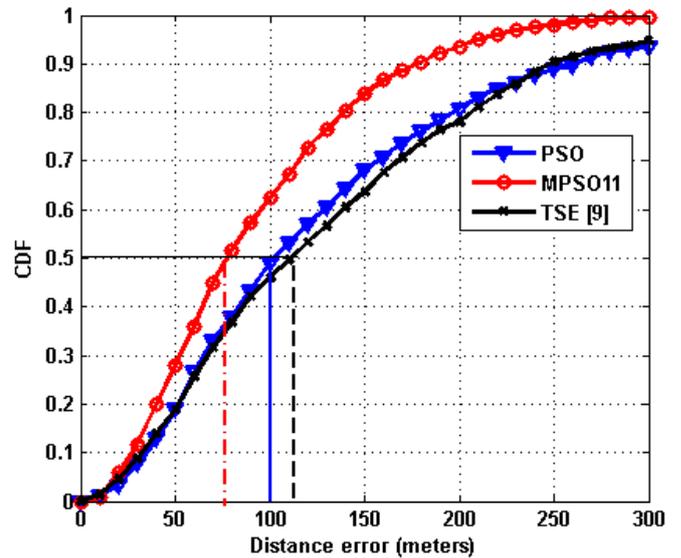

**Figure 5** The CDF for the thee algorithms namely PSO, MPSO11 and TSE for 10 SUs are cooperated together to localize the attacker that is located at (8000, 1000) inside the CR network and the SNR=-10dB

## 7. Conclusions

CRN is based on spectrum sensing mechanisms to define the free spaces in the spectrum left unused by primary users. An attacker can take advantage of this features by mimicking the primary user signals and performs PUEA and prevents the CRN form utilizing the free spaces. In this paper, the problem of PUEA is solved using cooperation between the CR users based on TDOA localization method, where the location of the PU is known as in the case of TV transmitters in WRAN 802.22 networks. The location of the emitter is measured and compared with the position of the TV tower to identify potential attacks. During the spectrum sensing process each SU records the signal and sends it to the CR base station, which derives the TDOA measurement by using cross-correlation methods. The PSO algorithms approaches are used to minimizing the nonlinear least squares cost, these approaches are developed by changing the inertia weights and acceleration coefficients of the original PSO. The best PSO approach is compared with the previous Taylor series method and the original PSO. The proposed approach obtained a low convergence time and high accuracy.

## Reference

[1] Mitola III, J., & Maguire Jr, G. Q. (1999). Cognitive radio: making software radios more personal. Personal Communications, IEEE, 6(4), 13-18.
[2] Yücek, T., & Arslan, H. (2009). A survey of spectrum sensing algorithms for cognitive radio applications .Communications Surveys & Tutorials, IEEE, 11(1), 116-130.

[3] Clancy, T. C., & Goergen, N. (2008, May). Security in cognitive radio networks: Threats and mitigation. In Cognitive Radio Oriented Wireless Networks and Communications, 2008. CrownCom 2008. 3rd International Conference on (pp. 1-8). IEEE.

[4] Chen, R., & Park, J. M. (2006, September). Ensuring trustworthy spectrum sensing in cognitive radio networks. In Networking Technologies for Software Defined Radio Networks, 2006. SDR'06.1 st IEEE Workshop on (pp. 110-119). IEEE.

[5] Ghanem, W. R., Essam, R., and Dessouky, M. (2018). Proposed Particle Swarm Optimization Approaches for Detection and Localization of the Primary User Emulation Attack in Cognitive Radio Networks. 35th National Radio Science Conference (NRSC2018), pp. 309–318

[6] Cordeiro, C., Challapali, K., Birru, D., & Sai Shankar, N. (2005, November). IEEE 802.22: the first worldwide wireless standard based on cognitive radios. In New Frontiers in Dynamic Spectrum Access Networks, 2005. DySPAN 2005. 2005 First IEEE International Symposium on (pp. 328-337). IEEE.

[7] W. R. Ghanem, M. Shokir, and MI Dessouky, "Investigation of puea in cognitive radio networks using energy detection in different channel model," Circuits and Systems: An International Journal, vol. 2, no. 2/3/4, pp. 11, 2015.

[8] Chen, C., Cheng, H., & Yao, Y. D. (2011). Cooperative spectrum sensing in cognitive radio networks in the presence of the primary user emulation attack. Wireless Communications, IEEE Transactions on, 10(7), 2135-2141.

[9] Blesa, J., Romero, E., Rozas, A., & Araujo, A. (2013). PUE attack detection in CWSNs using anomaly detection techniques. EURASIP Journal on Wireless Communications and Networking, 2013(1), 1-13.

[10] Le, T. N., Chin, W. L., & Kao, W. C. (2015). Cross-Layer Design for Primary User Emulation Attacks Detection in Mobile Cognitive Radio Networks. IEEE Communications Letters, 19(5), 799 - 802.

[11] W. R. Ghanem, M. Shokir, and M. Dessoky, "Defense against selfish PUEA in cognitive radio networks based on hash message authentication code," International Journal of Electronics and Information Engineering, vol. 4, no. 1, pp. 12–21, 2016.

[12] Chen, R., Park, J. M., & Reed, J. H. (2008). Defense against primary user emulation attacks in cognitive radio networks. Selected Areas in Communications, IEEE Journal on, 26(1), 25-37.

[13] León, O., Hernández-Serrano, J., & Soriano, M. (2012). Cooperative detection of primary user emulation attacks in CRNs. Computer Networks, 56(14), 3374-3384.

[14] Patwari, N., Ash, J. N., Kyperountas, S., Hero III, A. O., Moses, R. L., & Correal, N. S. (2005). Locating the nodes: cooperative localization in wireless sensor networks. Signal Processing Magazine, IEEE, 22(4), 54-69.

[15] Mao, G., Fidan, B., & Anderson, B. D. (2007). Wireless sensor network localization techniques. Computer networks, 51(10), 2529-2553.

[16] Fister Jr, I., Yang, X. S., Fister, I., Brest, J., & Fister, D. (2013). A brief review of nature-inspired algorithms for optimization. arXiv preprint arXiv:1307.4186.

[17] Gopakumar, A., & Jacob, L. (2008). Localization in wireless sensor networks using particle swarm optimization.

[18] W. R. Ghanem, M. Shokair and M. I. Desouky (Feb 22- 25,2016), "An improved primary user emulation attack detection in cognitive radio network based on Firefly Optimization Algorithm", 33rd National Radio Conference (NRSC 2016), pp. 178-187.

[19] Sivakumar, S., & Venkatesan, R. (2015). Meta-heuristic approaches for minimizing error in localization of wireless sensor networks. Applied Soft Computing, 36, 506-518.

[20] Monica, S., & Ferrari, G. (2014). Swarm intelligent approaches to auto-localization of nodes in static UWB networks. Applied Soft Computing, 25, 426-434.

[21] So, H. C. (2011). Source localization: algorithms and analysis. Handbook of Position Location: Theory, Practice, and Advances, 25-66.

[22] Kennedy, J., & Eberhart, R. (1995, November). Particle swarm optimization. In Neural Networks, 1995. Proceedings, IEEE International Conference on (Vol. 4, pp. 1942-1948). IEEE.

[23] Shi, Y., & Eberhart, R. (1998, May). A modified particle swarm optimizer. In Evolutionary Computation Proceedings, 1998. IEEE World Congress on Computational Intelligence, The 1998 IEEE International Conference on (pp. 69-73). IEEE.

[24] Xin, J., Chen, G., & Hai, Y. (2009, April). A particle swarm optimizer with multi-stage linearly-decreasing inertia weight. In Computational Sciences and Optimization, 2009. CSO 2009. International Joint Conference on (Vol. 1, pp. 505-508). IEEE.

[25] Eberhart, R. C., & Shi, Y. (2001). Tracking and optimizing dynamic systems with particle swarms. In Evolutionary Computation, 2001. Proceedings of the 2001 Congress on (Vol. 1, pp. 94-100). IEEE.

[26] Al-Hassan, W., Fayek, M. B., & Shaheen, S. I. (2006, November). Psosa: An optimized particle swarm technique for solving the urban planning problem. InComputer Engineering and Systems, The 2006 International Conference on (pp. 401-405). IEEE.

[27] Chen, G., Huang, X., Jia, J., & Min, Z. (2006, June). Natural exponential inertia weight strategy in particle swarm optimization. In Intelligent Control and Automation, 2006. WCICA 2006. The Sixth World Congress on (Vol. 1, pp. 3672-3675). IEEE.

[28] Kentzoglanakis, K., & Poole, M. (2009, July). Particle swarm optimization with an oscillating inertia weight. In Proceedings of the 11th Annual conference on Genetic and evolutionary computation (pp. 1749-1750). ACM.

[29] Feng, Y., Teng, G. F., Wang, A. X., & Yao, Y. M. (2007, September). Chaotic inertia weight in particle swarm optimization. In Innovative Computing, Information and Control, 2007. ICICIC'07. Second International Conference on(pp. 475-475). IEEE.

[30] Chatterjee, A., & Siarry, P. (2006). Nonlinear inertia weight variation for dynamic adaptation in particle swarm optimization. Computers & Operations Research, 33(3), 859-871.

[31] Lei, K., Qiu, Y., & He, Y. (2006, January). A new adaptive well-chosen inertia weight strategy to automatically harmonize global and local search ability in particle swarm optimization. In Systems and Control in Aerospace and Astronautics, 2006. ISSCAA 2006. 1st International Symposium on (pp. 4-pp). IEEE.

[32] Ting, T. O., Shi, Y., Cheng, S., & Lee, S. (2012). Exponential inertia weight for particle swarm optimization. In Advances in swarm intelligence (pp. 83-90). Springer Berlin Heidelberg.

[33] Nickabadi, A., Ebadzadeh, M. M., & Safabakhsh, R. (2011). A novel particle swarm optimization algorithm with adaptive inertia weight. Applied Soft Computing, 11(4), 3658-3670.

[34] Rezaee Jordehi, A., & Jasni, J. (2013). Parameter selection in particle swarm optimisation: a survey. Journal of Experimental & Theoretical Artificial Intelligence, 25(4), 527-542.

[35] Cui, Z., Zeng, J., & Yin, Y. (2008, November). An improved PSO with time-varying accelerator coefficients. In Intelligent Systems Design and Applications, 2008. ISDA'08. Eighth International Conference on (Vol. 2, pp. 638-643). IEEE.

[36] Mirjalili, S., Lewis, A., & Sadiq, A. S. (2014). Autonomous particles groups for particle swarm optimization. Arabian Journal for Science and Engineering, 39(6), 4683-4697.